\documentclass[a4paper,12pt]{article}
\usepackage{epsfig}
\usepackage{amssymb}

\begin{document}

\begin{titlepage}
\begin{center} \huge
  Photoproduction at HERA
\end{center}
\begin{center}
  \large J. Chwastowski and J. Figiel
\end{center}
\begin{center} \normalsize \it
  H. Niewodnicza\'nski Inst. of Nuclear Physics \\
  Polish Academy of Sciences,\\
  ul. Radzikowskiego 152,
  31-342 Cracow, Poland
\end{center}
\vspace*{1.cm}
%\centerline{Draft 3.141 \today }
\vspace*{1.cm}
\begin{center}
  \begin{minipage}{13.cm} \noindent 
\centerline{\bf Abstract} 
\vspace*{0.5cm}

Selected aspects of photoproduction in $ep$ scattering at the HERA collider, 
studied with the ZEUS detector, are presented.
The results are interpreted in the formalism of Vector Dominance Model, 
Regge theory and perturbative Quantum Chromodynamics.
\end{minipage} 
\end{center}
\end{titlepage}
 
\pagenumbering{roman}
\tableofcontents
\clearpage
% this gives wrong page number
%\addcontentsline{toc}{chapter}{List of Figures}
%\listoffigures 

% this gives wrong page number
%\addcontentsline{toc}{chapter}{List of Tables}
%\listoftables

\pagenumbering{arabic}
\section{Introduction}
\label{sec:intro}
\begin{figure}[ht]
\begin{center}
\epsfig{file=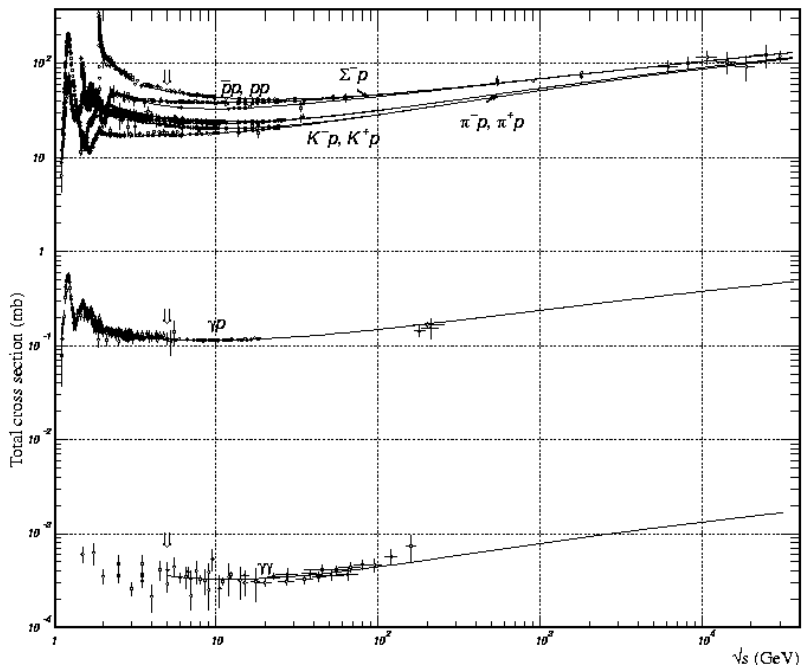,width=0.8\textwidth}
\end{center}
\caption{Comparison of the hadronic, $\gamma p$ and $\gamma\gamma$ total cross
sections as a function of the centre-of-mass energy. From \cite{pdg}.}
\label{fig:sigpdg}
\end{figure}

Experiments show that photoproduction on nucleons has features similar 
to hadron--hadron collisions \cite{bauer}. The energy dependence of the total
cross section resembles that of hadron--nucleon scattering 
(see Fig. \ref{fig:sigpdg}). For low energies a complicated structure 
corresponding to the formation of excited states or resonances is observed. 
Above about 3 GeV the cross section initially decreases and for larger 
centre-of-mass energies it increases slowly with energy. 
\begin{figure}[ht]
\begin{center}
\epsfig{file=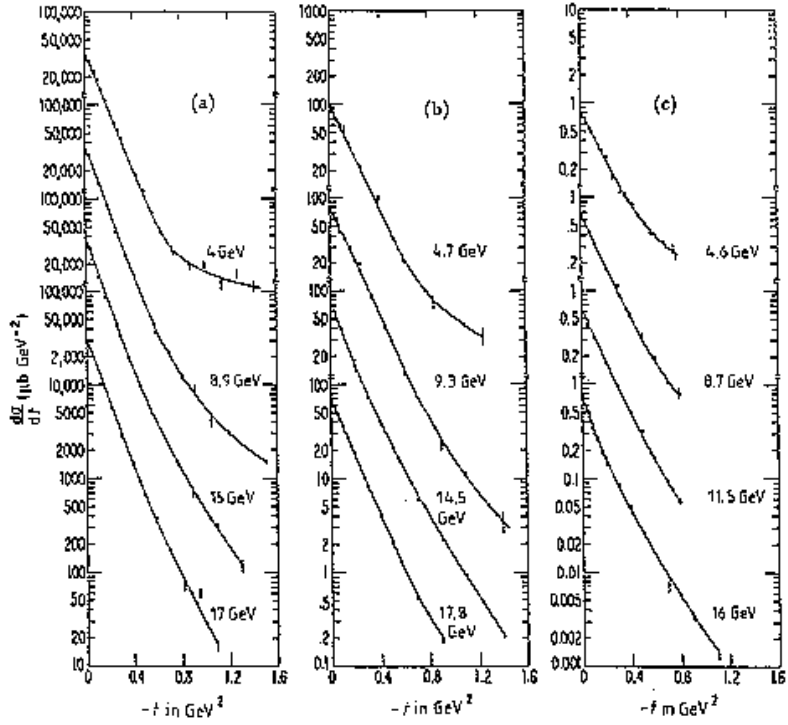,width=0.8\textwidth}
\end{center}
\caption{ Comparison of the elastic cross section $d\sigma/dt$ for three 
reactions: (a) $\pi^- p \rightarrow \pi^- p$, 
(b) $\gamma p \rightarrow \rho^0 p$, and (c) $\gamma p \rightarrow \gamma p$. 
From \cite{kogan}.}
\label{fig:diffpeak}
\end{figure}
Compton scattering, $\gamma p \rightarrow \gamma p$, shows a forward 
diffraction peak \cite{kogan} (see Fig. \ref{fig:diffpeak}) and its amplitude 
is predominantly imaginary \cite{alven}. As can be seen from 
Fig. \ref{fig:diffpeak} the elastic cross section, $d\sigma/dt$, for the three
reactions: $\pi^- p \rightarrow \pi^- p$, $\gamma p \rightarrow \rho^0 p$ and
$\gamma p \rightarrow \gamma p$, follows a similar behaviour with comparable 
values of the nuclear slope parameter, $b$. A copious production of the neutral
vector mesons is one of the most striking features of the photoproduction.\\
To first approximation a photon is an object with point-like interaction. 
However, it can quantum--mechanically fluctuate into a fermion--anti-fermion 
pair. The photon fluctuations into a pair of virtual charged leptons is 
described by QED. 
The photon can also fluctuate into a $q\bar{q}$ state with the photon quantum 
numbers ($J^{PC} = 1^{--}, Q = S = B =0$). Innteraction between the $q\bar{q}$ 
pair and the proton will occur if the fluctuation time \cite{ioffe}, $t_f$, is 
large compared to the interaction time, $t_i$. From the uncertainty principle 
the fluctuation time is given by
%\begin{equation}
$$
t_f = \frac{2E_{\gamma}}{m^2_{q\bar{q}}}
\label{eqn:ioffe}
$$
%\end{equation}
where $E_\gamma$ is the photon energy in the proton rest frame and 
$m_{q\bar{q}}$ is the mass of the $q\bar{q}$ state. The interaction
time is of the order of the proton radius $t_i \approx 1$ fermi.
For interactions of 10 GeV photons with a proton at rest, assuming that the
$q\bar{q}$-pair is the $\rho$ meson, $t_f \approx 7$ fermi so the condition 
$t_f \gg t_i$ holds. For a virtual photon the fluctuation time is
$$
t_f = \frac{2E_{\gamma}}{m^2_{q\bar{q}}+Q^2}
$$
where $Q^2$ is the photon virtuality. As $Q^2$ increases, the fluctuation time 
gets smaller (for fixed $m_{q\bar{q}}$) and the photon behaves more like a 
point-like object.\\
The above picture can be used for the photon--proton scattering subprocesses 
classification \cite{frisjo}. The scale of $q\bar{q}$ fluctuations can be
characterised by the transverse momentum $p_T$ of the $q\bar{q}$ system with
respect to the photon direction. Small scales result in long lived 
fluctuations, for which there is enough time to develop a gluon cloud around 
the $q\bar{q}$ pair. This is the domain of non-perturbative QCD physics. 
Usually photoproduction of such pairs is described by a sum over low mass 
vector states (the vector meson dominance model - VDM \cite{vdm}). The 
high-$p_T$ part should be perturbatively calculable. \\
Summarising, the photon can have three states: the ``point-like'' photon, 
the vector meson state and the perturbative $q\bar{q}$ pair \footnote{In the 
following the fluctuations into the charged lepton pair are neglected.}. 
This leads to three classes for $\gamma p$ interactions:
\begin{itemize}
\item the VDM class: a photon turns into a vector meson which subsequently 
interacts with the proton. This class contains all event types known from 
hadron induced reactions: elastic and diffractive scattering, a low- and 
high-p$_t$ non-diffractive interactions,
\item the direct class: photon undergoes a point-like interaction with a parton
                      from the proton,
\item the anomalous class: the photon perturbatively branches into a $q\bar{q}$
state and one of its partons interacts with a parton from the proton.
\end{itemize}
Experimentally the high-p$_t$ non-diffractive interaction of the VDM class and
the anomalous processes are joined into the resolved processes.

\section{Experimental Environment}
\label{sec:exper}
The HERA $ep$ storage ring  \cite{hera} is well suited to study 
photoproduction at high energies. The energy of the electron or positron beam 
is 27.5 GeV. The proton beam energy was increased to 920 GeV from 820 GeV in 
1998.\\
The results presented in the following were obtained by the ZEUS collaboration.
The collaboration operates a general purpose magnetic detector \cite{zeusblue}.
Charged particles are tracked in the central tracking detector (CTD) \cite{ctd}
which operates in a magnetic field of 1.43 T provided by a thin superconducting
solenoid. The high-resolution uranium-scintillator calorimeter \cite{ucal} 
(CAL) covers 99.7\% of the solid angle. It consists of three parts: the forward
(FCAL), the barrel (BCAL) and the rear (RCAL) calorimeters. Each CAL part is 
longitudinally segmented into electromagnetic and hadronic sections.
Each section is further subdivided transversely into cells.  
%The calorimeter
%gives equal response to electrons and hadrons (e/h = 1.00$\pm$0.02). 
Its relative energy resolution for electromagnetic showers is 
$0.18/\sqrt{E(GeV)}\oplus 0.01$ and for hadronic showers it is 
$0.35/\sqrt{E(GeV)}\oplus 0.02$ under test-beam conditions.
The HERA luminosity is measured via the rate of bremsstrahlung photons from the
Bethe--Heitler process emitted at angles $\Theta_{\gamma} \leq 0.5$ mrad. The 
photons are registered in a lead/scintillator sandwich calorimeter 
\cite{lumiexp}. It is screened from the synchrotron radiation by a carbon 
filter. The resulting relative energy resolution is about $0.23/\sqrt{E (GeV}$.
A typical systematic uncertainty on the luminosity measurement is 1-2\%.\\
A system of electron taggers consists of three calorimeters placed
at 8, 35 and 44 meters away from the nominal interaction point. They tag 
scattered electrons in a wide energy range. In addition lead/scintillator 
sandwich calorimeter placed at 35 meters is used to measure the scattered 
electron energy. Its relative energy resolution is about $0.20/\sqrt{E (GeV)}$.
The scattered electron energy range registered by this device is 
$5 \lesssim E_e^\prime \lesssim 20$ GeV. \\
In fixed target experiments the photoproduction was studied by observing events
induced by real photons. The photons were produced in the Bethe--Heitler
process occurring when an electron passed a radiator. The measurement of the
final state electron yielded the photon energy. At HERA the electron beam is a 
source of quasi-real photons and the photoproduction events are divided into 
two classes. In the first one, ``tagged events'' class, the final state 
electron is measured in the electron taggers. For such events the photon 
virtuality, $Q^2$, is restricted to $Q^2_0 < Q^2 < 0.02$ GeV$^2$ where 
$Q_0^2 = m_e^2y^2/(1-y)$ is the minimum value of $Q^2$ at a fixed value of the 
electron inelasticity $y$. The ``untagged events'' sample is defined requesting
that the final state electron is not observed in the CAL. This requirement 
limits the photon virtuality to  $Q^2 < 4$ GeV$^2$ with the median 
$Q^2 \approx 5\cdot 10^{-5}$ GeV$^2$.

\section{Photoproduction Total Cross Section}
\label{sec:sigtot}
The photon--proton total cross section was measured \cite{zeussigtot} in the 
process $e^+ p \rightarrow e^+\gamma p \rightarrow e^+ X$ with the ZEUS 
detector at HERA. The measurement was carried out for photons with virtuality
$Q^2 < 0.02$ GeV$^2$ and at the average photon--proton centre-of-mass energy 
$W_{\gamma p} = 209$ GeV. The data were collected in a dedicated run, 
to control systematic effects, with an integrated luminosity of 49 nb$^{-1}$.
The measured cross section is 
$\sigma_{TOT}^{\gamma p} = 174 \pm 1 (stat.) \pm 13(syst.)$.
\begin{figure}[ht]
\begin{center}
\epsfig{file=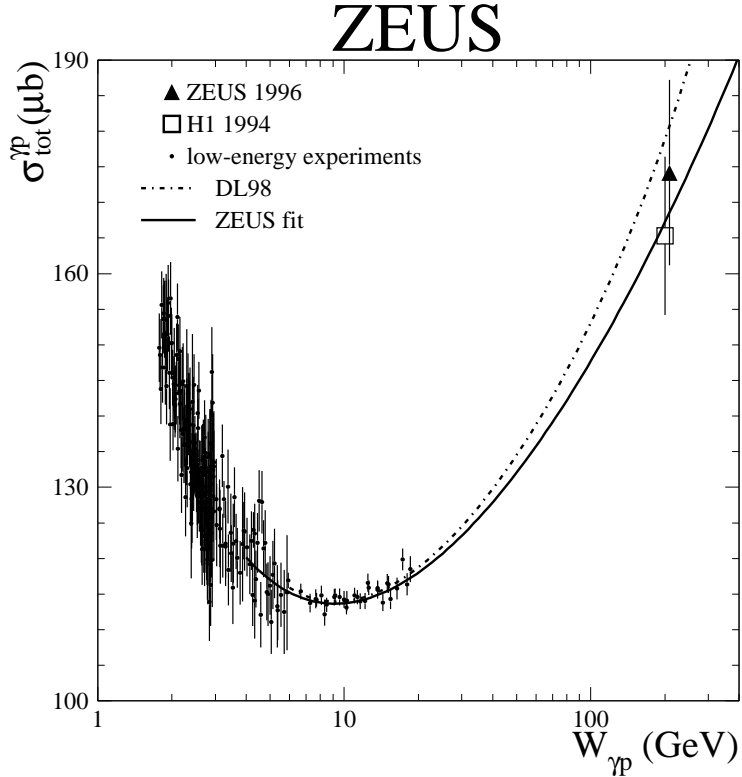,width=0.8\textwidth}
\end{center}
\caption{Photoproduction total cross section as a function of energy. ZEUS 
measurement (filled triangle), low energy data (filled circles), H1 measurement
(open square), the DL98 parameterisation (dash-dotted line) and the ZEUS fit 
(solid line). From \cite{zeussigtot}.}
\label{fig:sigtot}
\end{figure}
The total photoproduction cross section as a function of energy is shown in
Fig. \ref{fig:sigtot}. The ZEUS result is in good agreement with H1 measurement
\cite{h1sigtot} at a similar centre-of-mass energy. Also the low energy data 
are shown in the figure. In addition the ZEUS collaboration extrapolated the 
measurements \cite{bpc} at low $Q^2$, $ 0.11 < Q^2 < 0.65$ GeV$^2$ to $Q^2 = 0$
using generalized VDM \cite{bpcgvdm}. The extrapolation yielded 
$\sigma^{\gamma p}_{TOT}= 187 \pm 5(stat.)\pm 14(syst.) \mu b$ at 
$W_{\gamma p} = 212$ GeV, a value which is somewhat larger but compatible with 
the direct measurement within errors.\\
A Regge theory \cite{regge} (see also \cite{collins}) motivated fit of the 
cross section energy dependence
%\begin{equation}
$$
\sigma_{TOT}^{\gamma p} = A\cdot W^{2\epsilon}_{\gamma p} 
                        + B\cdot  W^{2\eta}_{\gamma p},
\label{eq:fit}
$$
%\end{equation}
similar to the one postulated in \cite{dl98} or \cite{cudell}, is shown in Fig.
\ref{fig:sigtot} as a solid line. The first term related to the pomeron 
intercept as $\alpha_{{I\!\!P}\/\/ }(0) = 1+\epsilon$ describes the high energy
behaviour of the cross section. The second term corresponds to the Reggeon 
exchange with the intercept $\alpha_{{I\!\!R}\/\/ }(0) = 1-\eta$. The fit was 
performed to all the $\gamma p$ data \cite{compass} with $W_{\gamma p} > 4$ GeV with reggeon intercept fixed to the value obtained by Cuddel et al. 
\cite{cudell} $ \eta = 0.358\pm0.015$. The fit yielded
$$
A = 57 \pm 5 \mu b; \hspace{1cm} B = 121 \pm 13 \mu b,
$$
and 
$$
 \epsilon = 0.100 \pm 0.012.
$$
The resulting value of $\epsilon$ is in good agreement with 
$\epsilon = 0.093\pm0.002$  obtained in \cite{cudell} from the analysis of 
hadronic data.\\
The Donnachie--Landshoff parameterisation \cite{dl98}, shown as a dash-dotted 
line in Fig. \ref{fig:sigtot} includes soft- and hard-pomeron trajectories. It
agrees with the ZEUS measurement within the errors. Also other 
parameterisations %based on the hadron -- hadron cross sections 
\cite{frisjo2,block,allm97} reproduce the ZEUS result.\\
In addition the total $\gamma \gamma$ cross section calculated from the 
assumption of the cross section factorisation 
$$
\sigma_{TOT}^{\gamma\gamma} \cdot \sigma_{TOT}^{pp}=(\sigma_{TOT}^{\gamma p})^2
$$
%and parameterisations of the $\sigma_{TOT}^{\gamma p}$ \cite{zeussigtot} 
%(outlined above) and the $\sigma^{pp}_{TOT}$ \cite{cudell} 
agrees well with LEP measurements \cite{l3fact,opalfact}.

\section{Elastic Vector Meson Production}
\label{sec:elastic}
\begin{figure}[ht]
\begin{center}
\epsfig{file=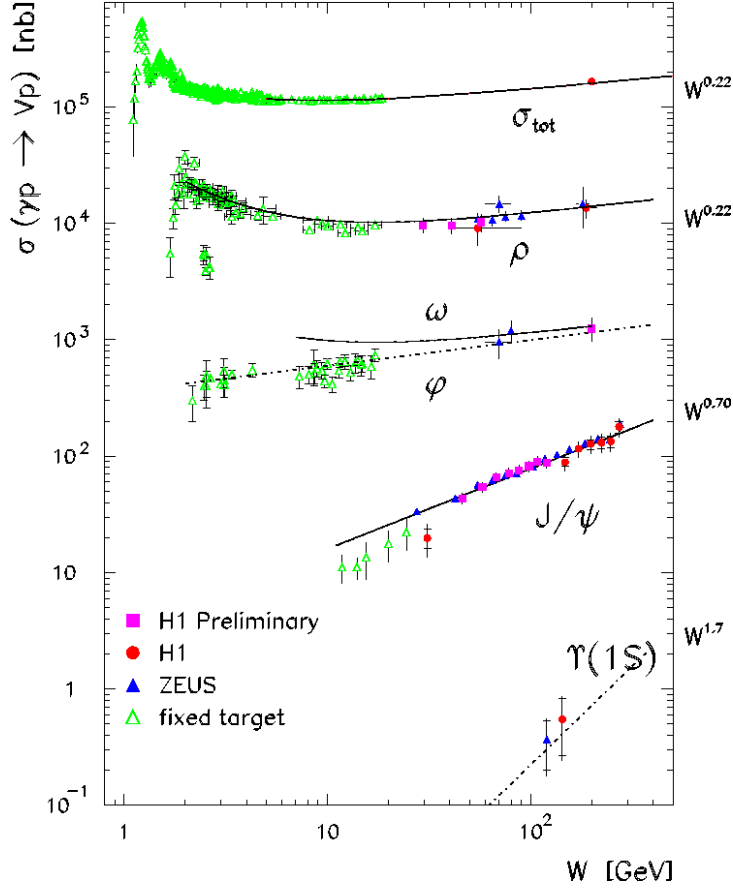,width=0.7\textwidth}
\end{center}
\caption{The total photoproduction cross section and the cross sections for 
elastic vector meson production ($\rho$, $\omega$, $\phi$ $J/\psi$ and 
$\Upsilon$) as a function of $W$. Lines show a $W^\delta$ 
dependence with $\delta$ values indicated. From \cite{vmvsw}.}
\label{fig:vmvsw}
\end{figure}

Elastic vector meson production is the process
$$
\gamma p \rightarrow V p
$$
where $V$ denotes one of the vector mesons. This reaction which was extensively
studied with real and virtual photons for photon--proton centre-of-mass energy,
$W$, below 20 GeV, exhibits features which are also characteristic for hadronic
diffractive reactions. The cross section energy dependence is weak and the 
dependence on $t$, the square of the  four-momentum transfer at the proton 
vertex, is approximately exponential i.e. $d\sigma/dt \sim e^{-b|t|}$. This 
similarity can be explained on the grounds of the VDM where the photon 
fluctuates into a long lived vector meson state and subsequently scatters on 
the proton. Regge theory predicts that at high energies the centre-of-mass
energy dependence of the cross section for $\rho$, $\omega$ and $\phi$ 
production is 
$$
\sigma_{\gamma p \rightarrow V p} \approx \frac{W^\delta}{b(W)}.
$$
%the effective exponent is 
%$$
%\delta \approx 4(\alpha_{{I\!\!P}\/\/ }(0)-\frac{\alpha^\prime_{{I\!\!P}\/\/ }}{b_0}-1).
%$$
%Using $\alpha_{{I\!\!P}\/\/ }(0)= 1.08$ as given by the Donnachie -- Landshoff 
%parametrisation 
%\cite{dl98,dl92} and $b_0 \approx 10$ results $\delta \approx 0.22$.
The energy dependence of the cross sections for elastic vector meson production
is shown in Fig. \ref{fig:vmvsw} together with the HERA measurements 
\cite{RHOMEAS,OMEGAMEAS,PHIMEAS,h1rho}. Also the data on the total 
photoproduction cross section are presented. The total cross section and that 
for the production of the lowest lying vector mesons show a similar dependence 
with $\delta \approx 0.22$.

\begin{figure}[ht]
\begin{center}
\epsfig{file=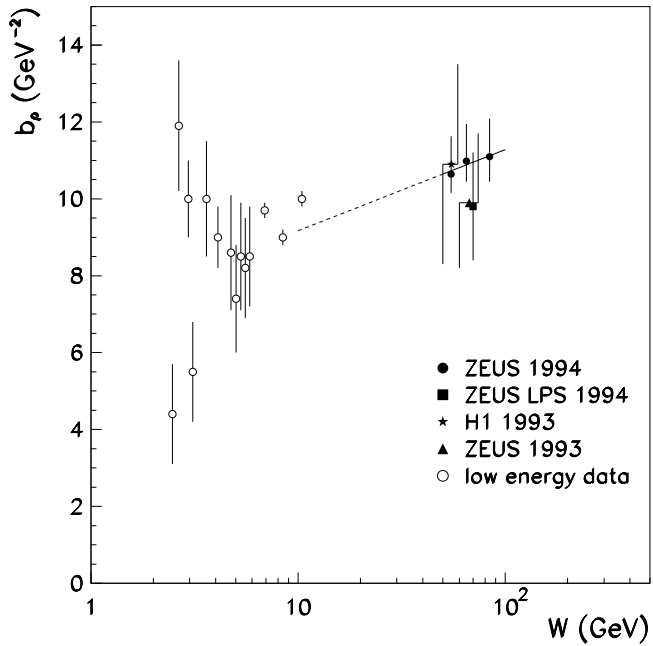,width=0.7\textwidth}
\end{center}
\caption{Compilation of the low energy \cite{lowbvsw} and HERA 
\cite{RHOMEAS,h1rho,herabvsw} results on the exponential slope parameter, $b$, 
for the elastic reaction $\gamma p \rightarrow \rho p$.
%in the kinematic region $50 < W <  100$ GeV and $|t| < 0.5$ GeV$^2$ 
%as a function of $W$. The low energy \cite{lowbvsw} and HERA 
%\cite{RHOMEAS,h1rho,herabvsw} results are presented. 
The solid line represents the fit of the energy dependence. The extrapolation 
of the fit to lower energies is marked by the broken line. 
From \cite{RHOMEAS}.}
\label{bvsw}
\end{figure}

For a linear pomeron trajectory the Regge prediction for the slope parameter, 
$b(W)$, is
$$
b(W) = b_0+2\alpha^\prime_{{I\!\!P}\/\/ } ln\frac{W^2}{W^2_0},
$$
where $\alpha^\prime_{{I\!\!P}\/\/ }$ is the pomeron trajectory slope and
$b_0$ and $W_0$ are constants. Figure \ref{bvsw} shows a compilation of the 
HERA \cite{RHOMEAS,h1rho,herabvsw} and low energy data \cite{lowbvsw} on the 
slope, $b$, in the case of the elastic reaction $\gamma p \rightarrow\rho p$. 
The Regge predictions are also depicted. The value of $b$ rises with increasing
$W$ suggesting the shrinkage of the $t$-distribution forward peak with energy. 
The growth of $b$ with photon--proton centre-of-mass energy is compatible with 
the Regge prediction.
% = b_0+2\alpha^\prime_{{I\!\!P}\/\/ } ln\frac{W^2}{W^2_0}$ 
%in the case of the elastic reaction $\gamma p \rightarrow\rho p$ is shown in 
%Fig. \ref{bvsw}. 
The ZEUS collaboration finds 
$\alpha^\prime_{{I\!\!P}\/\/ } = 0.23 \pm 0.15(stat.)^{+0.10}_{-0.07}(syst.)$ 
GeV$^{-2}$ consistent with the value of 0.25 GeV$^{-2}$ obtained \cite{dl92} 
from hadron--hadron elastic scattering.

The photoproduction of $J/\Psi$ was measured \cite{h1psi1,h1psi2,zeuspsi} at 
HERA and is shown in  Fig. \ref{fig:vmvsw}. The  $J/\Psi$ photoproduction cross
section has much stronger energy dependence with power $\delta \approx 0.7$.
This behaviour can be explained by perturbative QCD in which the pomeron is 
interpreted as a two-gluon colour singlet exchange. In pQCD the steep increase 
of the cross section is connected \cite{ryskin} with the rise of the gluon 
density in the proton with decreasing $x$ (increasing $W$). 
%However, the normalisation is not predicted correctly \cite{ryskin2}. 
Perturbative QCD states that the cross section is proportional to the square of
the gluon density function of the proton, i.e.
$$
 \sigma \sim [\hat{x}g(\hat{x},\hat{q}^2)]^2
$$
with $\hat{q}^2 = (Q^2+m^2_{J/\Psi}+|t|)/4$ and
$\hat{x} = (Q^2+m^2_{J/\Psi}+|t|)/W^2$. The mass of the $J/\Psi$ meson
provides a scale large enough, $\hat{q}^2 \simeq 2.5$ GeV$^2$, for the 
perturbative QCD calculations to be valid. \\
Both HERA collaborations extracted the effective pomeron trajectory from the 
energy dependence of the slope parameter. The fitted pomeron trajectories are 
compatible within the errors and H1 measures \cite{h1psi2} 
$\alpha_{{I\!\!P}\/\/ }(0) = (1.20\pm0.02)$, 
$\alpha^\prime_{{I\!\!P}\/\/ }=(0.15\pm0.06)$ GeV$^{-2}$
while ZEUS \cite{zeuspsi} 
$\alpha_{{I\!\!P}\/\/ }(0) = (1.200\pm0.009)$, 
$\alpha^\prime_{{I\!\!P}\/\/ }=(0.115\pm0.018(stat.)^{+0.008}_{-0.015}(syst.))$
 GeV$^{-2}$
in a similar kinematic range. The soft-pomeron trajectory 
$\alpha_{{I\!\!P}\/\/ }(t) = 1.08 + 0.25\cdot |t|$ is inconsistent with the 
above findings.  \\
Measurements of the decay angular distribution of vector mesons photoproduced 
at small four-momentum transfer show that they have the same helicity as the 
interacting photon.
This fact is called the s-channel helicity conservation (SCHC) and is typical
for soft diffractive processes.\\
The elastic photoproduction of the $\Upsilon$ meson was measured 
\cite{h1psi1,zeusupsi} via its decay into a $\mu^+\mu^-$ pair. No distinction 
for $\Upsilon$, $\Upsilon^\prime$ and $\Upsilon''$ was made due to the limited 
experimental resolution. The cross section is small and below 1 nb (see Fig. 
\ref{fig:vmvsw}). 
%Its energy dependence is even stronger than the one observedfor $J/\Psi$. 
The $\Upsilon$ photoproduction cross section was found to be 
reasonably well described by the pQCD calculations. These calculations are 
either based on the leading vector meson cross section including corrections 
\cite{frankfurtupsi} or use the parton hadron duality hypothesis to obtain the 
production of $\Upsilon$ from the $b\bar{b}$ cross sections \cite{martinupsi}.

\section{Proton-dissociative Vector Meson Production}
\label{sec:prodiff}
\par
\begin{figure}[h]
\vspace*{-0.5cm}
\begin{center}
\epsfig{file=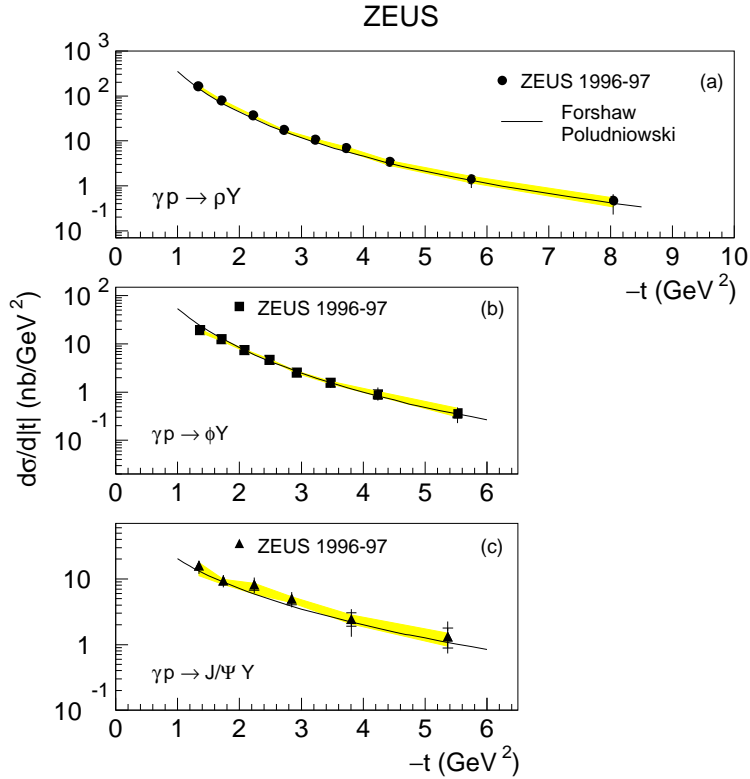,width=0.8\textwidth}
\end{center}
\vspace*{-0.5cm}
\caption{The $t$ distributions for $\rho$, $\phi$ and $J/\psi$ mesons in
proton-dissociative photoproduction. The shaded bands represent uncertainties
due to the modeling of hadronic system Y and the lines - the pQCD calculations 
described in the text. From \cite{zeusvmt2}.
\label{fig:larget1} }
\end{figure}

The ZEUS collaboration measured the proton-dissociative (double diffractive) 
photoproduction of the vector mesons: 
$$ \gamma p \rightarrow V Y $$
where $Y$ denotes system in which the proton dissociates diffractively. The
trigger conditions and the selection cuts ensured presence of a large rapidity 
gap ($\Delta\eta > 2$) between the vector meson and the system $Y$. The 
variable $\eta$ is the pseudorapidity of a particle defined as 
$\eta = 0.5\cdot log(tan(\Theta/2))$ where $\Theta$ is the particle polar angle
measured with respect to the proton direction.  The production of $\rho$, 
$\phi$ and $J/\psi$ mesons, in a large $|t|$ range: $1.2 < |t| < 10$ GeV$^2$, 
at a photon--proton centre-of-mass energy $80 < W < 120$ GeV and $Q^2 < 0.02$ 
GeV$^2$ was studied \cite{zeusvmt2}. In contrast to the elastic vector meson 
production and in accord with perturbative QCD expectation the differential
cross section follows a power law dependence $|t|^{-n}$, the heavier the meson 
the softer the distribution: $n = 3.21\pm0.04\pm0.15$ for $\rho$, 
$n = 2.7\pm0.1\pm0.2$ for $\phi$ and $n = 1.7\pm0.2\pm0.2$ for $J/\psi$. These 
data were successfully described by the pQCD calculations of \cite{polud}. The 
comparison of the data and calculations is depicted in Figure 
\ref{fig:larget1}. In this model, the virtual photon fluctuates into a 
$q\bar{q}$ dipole which couples via a gluon ladder (the BFKL pomeron) to a 
single parton ($\approx $ gluon) in the proton and then recombines into a 
vector meson. A non-relativistic approximation of the vector meson wave 
function was used (which is very approximate in the case of light mesons). 
With three parameters fitted to the data this model reproduces nicely both the 
shapes and the normalisation for the three vector mesons. On the other hand the
two-gluon exchange failed to describe these data. Additionally, the two-gluon 
exchange predicts an energy independent cross section, contrary to the BFKL 
pomeron exchange, which foresees its rise \cite{Enberg}. The recent results on 
$J/\psi$ photoproduction at large $t$ by the H1 collaboration \cite{H103} 
support this expectation and indicate the BFKL nature of the QCD pomeron in 
these processes.
\\
The simultaneous measurement of the $W$ and $t$ dependence allowed a
determination of the pomeron trajectory slope: 
$\alpha^\prime = -0.02\pm0.05(stat.)^{+0.04}_{-0.08}(syst.)$ and 
$\alpha^\prime = -0.06\pm0.12(stat.)^{+0.05}_{-0.09}(syst.)$ for the $\rho$ 
and $\phi$ meson, respectively. These values are in agreement with the pQCD 
expectations \cite{pqcdvmt} and are smaller than $\alpha^\prime =0.25$ 
GeV$^{-2}$ 
characteristic for soft processes at $-t < 0.5$ GeV$^2$ and also than those
measured for $-t < 1.5$ GeV$^2$ \cite{zeusvmt1}. These observations establish 
$|t|$ as a hard pQCD scale similarly to $Q^2$ in DIS. More quantitative 
comparison can be obtained by plotting the ratios of vector meson cross 
sections in function of both scales, $Q^2$ and $|t|$ \cite{zeusvmt2}. Under
simplifying assumptions that the photon couples directly to quarks in the 
vector meson and that the coupling does not depend neither on the vector meson 
mass nor its wave function (which seems reasonable in hard scattering) these
ratios reach SU(4) values of 2/9 for $\phi/\rho$ and 8/9 for $J/\Psi/\rho$. In 
fact the $\phi/\rho$ ratios approach the SU(4) values with increasing $Q^2$ and
$|t|$, as well as $\Psi/\rho$ in photoproduction ($|t|$). Generally, however 
the cross sections ratios rise faster with increasing $|t|$ than with $Q^2$ so 
these scales seem not to be  equivalent.
\\
The analysis of the angular distributions of the meson decay products was used
to determine the $\rho$ and $\phi$ spin-density matrix elements 
\cite{zeusvmt2}. They are $r^{04}_{00}$ and $r^{04}_{10}$ related to the single
helicity flip amplitudes and $r^{04}_{1-1}$ related to the double helicity 
flip amplitudes. Following the pQCD predictions and contrary to soft 
diffractive processes in which the helicity is conserved (SCHC hypothesis), all
these matrix elements are significantly different from zero: $r^{04}_{00}$ and 
${\cal R}e(r^{04}_{10})$ $\approx 0.05$ and $r^{04}_{1-1} \approx -0.15$ in the
whole $|t|$-range considered \cite{zeusvmt2}. These observations are
semi-quantitatively reproduced in a BFKL framework \cite{EnbergAPP}.
\section{Inclusive Diffraction}
\label{sec:diff}
A photon can dissociate not only into vector mesons but also into a 
multiparticle hadronic state $(X)$, of mass $M_X$ in the process of inclusive 
diffraction:
$$ \gamma p \rightarrow Xp $$
if the coherence condition $M_{X}^{2}/W^2 \ll 1 $ is satisfied. The E-612 
experiment at Fermilab studied this reaction in scattering of real photons off 
protons in the kinematic range $75 < E_\gamma < 148$ GeV, $0.02 < |t| < 0.1$ 
GeV$^2$ and $M_X^2/W^2<0.1$. At low mass it was found that the cross section is
dominated by the $\rho$ production.  The $t$-distribution in the $\rho$ mass 
region is exponential with a slope parameter $b = 10.6\pm1.0$ GeV$^{-2}$. For
larger masses the slope of the $t$-distribution is roughly half of that for the
$\rho$ region. At high values of $M_X^2$ a dominant $1/M_X^2$ behaviour 
was observed. 
%This behaviour is consistent with a large triple-pomeron presence
%in the diffractive amplitude. 
The diffractive events were characterised by the lack of hadronic
activity between the photon system $X$ and the final proton. This topological 
feature of the diffractive final state is called Large Rapidity Gap (LRG).
%The diffractive events show a characteristic topological feature. There is a 
%lack of hadronic activity between the photon system $X$ and the final proton.
\\
Photon inclusive diffractive dissociation was studied by H1 \cite{h1gdiss} and 
ZEUS \cite{dissfrac} collaborations using the LRG signature. Experimentally, 
the events were selected requiring a pseudorapidity gap $\Delta\eta$ between 
the most forward hadron (of pseudorapidity $\eta_{max}$) and the final proton.
The H1 collaboration carried out measurements at $W = 187$ GeV and $W = 231$ 
GeV. They found that the energy and $M_X^2$ dependence of the H1 data and the 
low energy data \cite{e612} is well described by the triple-Regge mechanism. 
The extracted effective intercept of the pomeron trajectory is 
$\alpha_{{I\!\!P}\/\/ }(0) = 
1.068 \pm 0.016(stat.) \pm 0.022(syst.) \pm 0.041(model)$
and agrees well with the one obtained for hadron--hadron scattering.\\
The ZEUS collaboration performed a study of the $M_X^2$ distribution at 
$W \approx 200$ GeV and found that for large masses ($8 < M_X^2 < 24$ GeV$^2$) 
the triple-Regge mechanism provides a good description of the data. The 
analysis yielded a value of the effective intercept of the pomeron trajectory
$\alpha_{{I\!\!P}\/\/ }(0)  = 1.12 \pm 0.04(stat.) \pm 0.07(syst.)$ which is
consistent with the one found by H1 within the errors.
The ZEUS collaboration found also that the ratio of the cross section for the 
photon diffractive dissociation to the total photoproduction cross section is 
$(13.3\pm 0.5(stat.)\pm 3.6(syst.))\%$.\\
\begin{figure}[ht]
\begin{center}
\epsfig{file=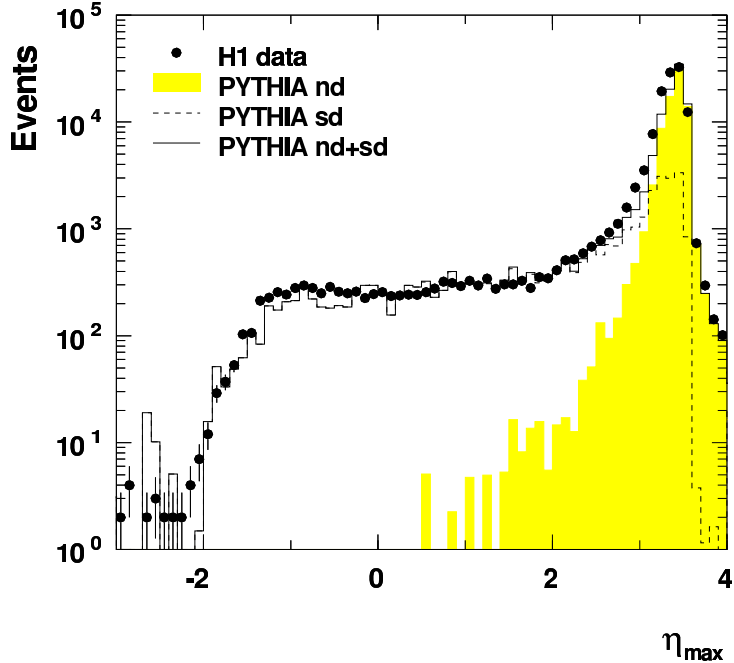,width=0.7\textwidth}
\caption{The $\eta_{max}$ distribution for photoproduction events containing 
jets of $E_T>5$ GeV and $-1.5 < \eta_{jet} < 2.5$ compared to the Pythia MC
predictions. The non-diffractive Pythia version is depicted by the shaded 
histogram, the diffractive one is marked by the dashed line and a sum of 
both by the solid line.
From \cite{h1jetlrg}.}
\label{h1jetlrg}
\end{center}
\end{figure}

In photoproduction the LRG signature is also observed in events with production
of jets. The large rapidity gap can be between the jets and the target particle
as proposed in the Ingelman--Schlein model \cite{islrg}. In this case the 
four-momentum transfer is small and target particle preserves its identity.
Bjorken \cite{bjorkenlrg} proposed to study the events in which the gap 
separates the jets. In such events the four-momentum transfer is large.\\
Both ZEUS and H1 analysed events with production of jets and a LRG in the 
proton fragmentation region \cite{zeusjetlrg,h1jetlrg}. Figure \ref{h1jetlrg} 
shows the $\eta_{max}$ distribution for such events. A clear excess of data 
over the non-diffractive Monte Carlo is observed for $\eta_{max} < 2$. The sum 
of the non-diffractive and diffractive Pythia MC \cite{pythia} well describes 
the data. \\
The ZEUS collaboration estimated in \cite{zeusjetlrg} that the gluon content of
the pomeron should be 30-80\% to describe the data with help of the 
Ingelman--Schlein model. \\
The measured cross sections for the 
diffractive dijet photoproduction \cite{h1diff2j,zeusdiff2j} show a steep 
fall-off with the transverse jet energy, $E_T^{jet}$, as expected for 
parton--parton scattering. \\
Recently, the H1 collaboration published an analysis \cite{h1diff2jb} of the 
diffractively produced jets in tagged photoproduction. They found that the 
Monte Carlo prediction, based on the H1 2002 QCD fit, well describes the shapes
of the differential cross sections. However, the normalisation is overestimated
by a factor of about 1.3. The shape of the differential cross section is well 
represented if in the Monte Carlo model the pomeron intercept of 
$\alpha_{{I\!\!P}\/\/ }(0)  = 1.17$ or $\alpha_{{I\!\!P}\/\/ }(0)  = 1.08$ is 
used while the choice of $\alpha_{{I\!\!P}\/\/ }(0)  = 1.4$ is disfavoured.\\
\begin{figure}[ht]
\begin{center}
\epsfig{file=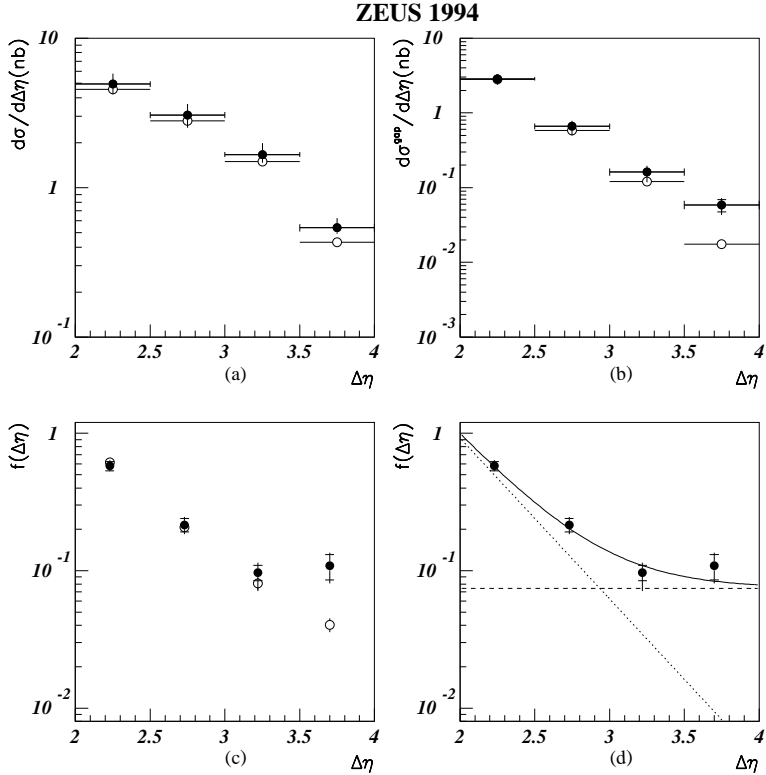,width=10cm}
\caption{Inclusive cross section, $d\sigma/d\Delta\eta$, as a function of the 
pseudorapidity distance $\Delta\eta$ between the jets (a) and for events with
the LRG signature (b).
ZEUS data - black circles. The Pythia prediction for non-singlet exchange - 
open circles. The gap fraction as a function of $\Delta\eta$ is depicted in (c)
and in (d) where also result of the fit (solid line) to a sum of an exponential
and a constant (dotted lines) is shown. From \cite{zeusjlrgj}.}
\label{fig:zjlrgj}
\end{center}
\end{figure}
Events with large rapidity gap between the jets were studied by the ZEUS 
\cite{zeusjlrgj} and H1 \cite{h1jlrgj} collaborations. Such events can be due 
to the exchange of a colour singlet object. The exchange of an electroweak 
boson or a strongly interacting colour singlet is possible. These exchanges 
would lead to similar results however their rates can be different. If the 
jets have large transverse energies then the four-momentum transfer is large 
and the process can be perturbatively calculated. Bjorken \cite{bjorkenlrg} 
estimated that the ratio of the colour singlet two-gluon exchange to the the 
single gluon exchange is about 0.1.\\
For events containing two jets the gap fraction, $f(\Delta\eta)$, is defined as
a number of dijet events with a certain gap size $\Delta\eta$ to the total 
number of dijet events for which the distance between the jets is $\Delta\eta$.
The ZEUS collaboration used events with at least two jets of $E_T >6$ GeV and 
separated in pseudorapidity by at least 2 units.
% so the jets do not overlap in $\eta$. 
The region between the jet cones with no particle of the transverse energy,
$E_T^{part.} > 250$ MeV is called a gap. The data and the Pythia MC predictions
are compared in Fig. \ref{fig:zjlrgj}. A clear excess of events for large 
values of $\Delta\eta$ is observed. The gap fraction for the colour singlet is 
found to be about $0.07 \pm 0.02 ^{+0.01}_{-0.02}$. It is larger than the 
values measured at Tevatron $\sim 0.01$ \cite{cdfgap,d0gap}. The H1 
collaboration used events with at least two jets with transverse energies 
$E_T^{j1} > 6$ GeV and $E_T^{j2} > 5$ GeV and separated by at least 2.5 
pseudorapidity units. In addition, H1  measures the total activity between the 
jets as $E_T^{gap}$ which is the sum of the transverse energies observed in the
region between two highest $E_T$ jets. For the lowest value of 
$E_T^{gap} < 0.5$ GeV and $3.5 < \Delta\eta < 4.0$ the gap fraction is 
approximately 10\% in good agreement with the ZEUS result.

\section{Hard Jets in Photoproduction}
\label{sec:hard}
\begin{figure}[ht]
\begin{center}
\epsfig{file=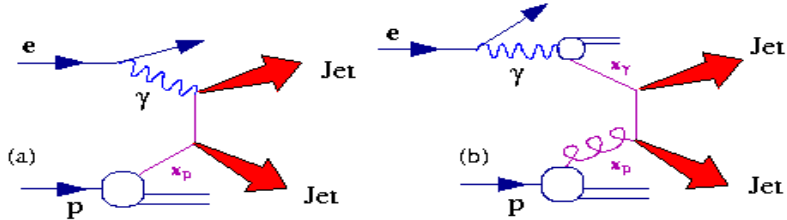,width=\textwidth,height=4cm}
\caption{Examples of the LO QCD diagrams for inclusive jet photoproduction
in direct (a) and resolved (b) processes.}
\label{fig:dijexmple}  
\end{center}
\end{figure}
\begin{figure}[th]
\begin{center}
\epsfig{file=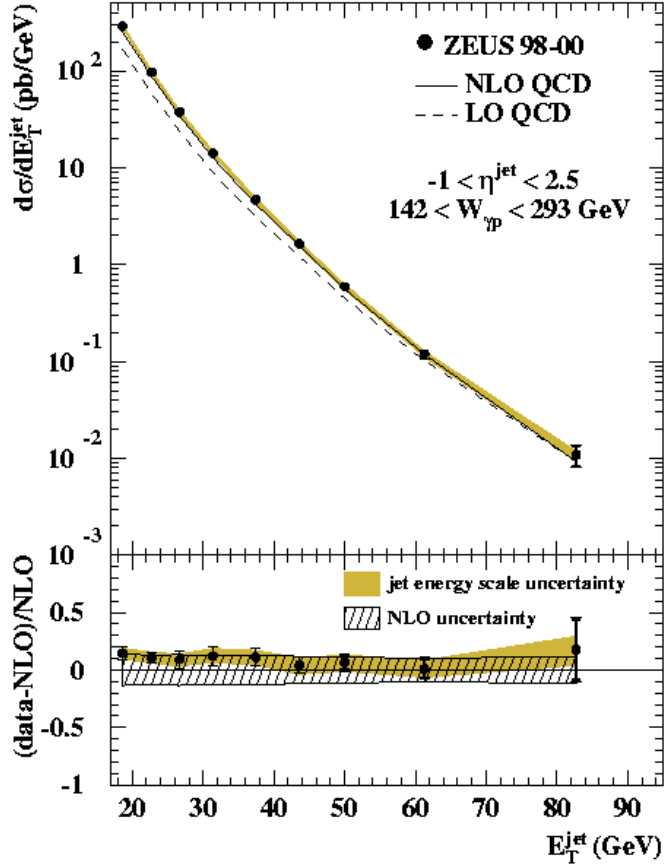,width=0.8\textwidth}
\caption{a) Measured inclusive jet cross section, $d\sigma/dE^{jet}_T$
(filled dots) compared to LO and NLO QCD calculations. The thick error bars 
represent the statistical uncertainties, the thin error bars show the 
statistical and systematic uncertainties added in quadrature. The shaded band 
shows the uncertainty associated to the absolute energy scale of the jets. The 
LO (dashed line) and NLO (solid line) QCD parton-level calculations corrected 
for hadronisation effects are also shown.
b) The fractional difference between the measured $d\sigma/dE^{jet}_T$ and the 
NLO QCD calculation with the calculation uncertainty marked by hatched band.
From \cite{zeusincljet}.}
\label{fig:zeusinclusivejet}  
\end{center}
\end{figure}
\begin{figure}[th]
\begin{center}
\epsfig{file=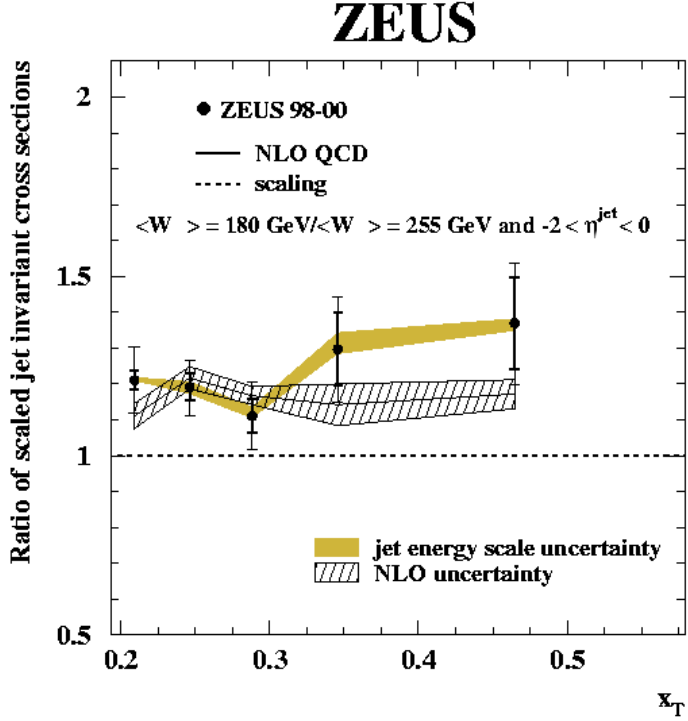,width=0.8\textwidth}
\caption{Measured ratio of the scaled jet invariant cross sections at
two $W$ intervals as a function of $x_T$. From \cite{zeusincljet}.}
\label{fig:zeusscaling}
\end{center}
\end{figure}
\begin{figure}[th]
\begin{center}
\epsfig{file=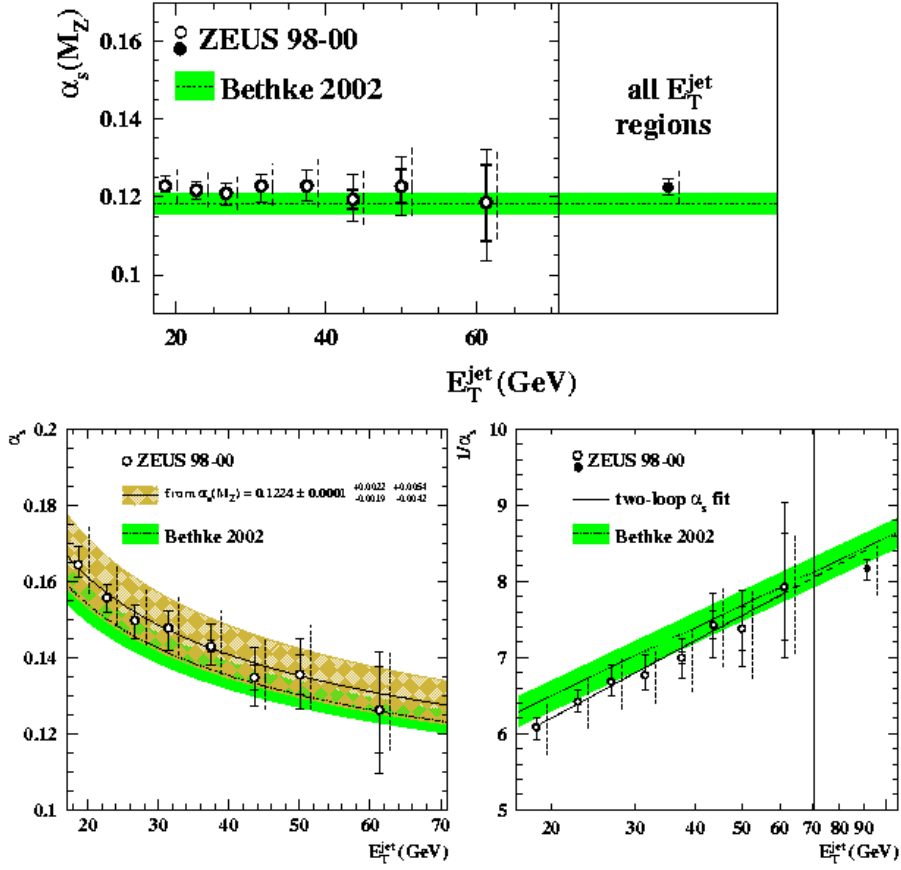,width=0.9\textwidth}
\caption{ a) The $\alpha_s(M_Z)$ values (open circles) 
as a function of $E_T^{jet}$. The combined result using all the $E_T^{jet}$ 
intervals is shown as a solid circle.
b) The value $\alpha_s(E_T^{jet})$ as a function of $E_T^{jet}$ (open circles).
The solid line represents the predictions for the central value of  
$\alpha_s(M_Z)$ measured by the ZEUS collaboration with the uncertainty given 
by the light-shaded band.
c) The value $1/\alpha_s(E_T^{jet})$ as a function of $E_T^{jet}$ (open 
circles).
The solid line represents the result of the two-loop $\alpha_s$ fit to the 
measured values. The dashed line shows the extrapolation to $E_T^{jet} = M_Z$.
In all figures the inner error bars show the statistical uncertainty and outer 
error bars represent the statistical and systematic uncertainties added in 
quadrature. The dashed error bars show the theoretical uncertainties.
The world average (dotted line) and its uncertainty (shaded band) are 
displayed. From \cite{zeusincljet}.}
\label{fig:zeusalfas}
\end{center}
\end{figure}

The photoproduction of jets at a large scale provided by the transverse energy,
$E_T$, of jets can be computed in perturbative QCD.  Examples of the leading
order QCD (LO QCD) diagrams for inclusive jet production are shown in Fig. 
\ref{fig:dijexmple}. In LO QCD such processes are divided into two classes. In 
the first one, called resolved process, the photon acts as source of partons 
and only a fraction of its momentum, $x_\gamma$, participates in the 
scattering. In the second one, the direct process, the photon interacts via 
boson-gluon fusion or QCD Compton scattering and acts as a point-like particle 
with $x_\gamma \approx 1$. Both classes lead to the production of jets. 
However, they differ in the jet topology. The resolved events contain the 
so-called photon remnant jet (see Fig. \ref{fig:dijexmple}). Jet cross sections
are sensitive to the photon and the proton structures and to the dynamics of 
the hard sub-process. For high $E_T$ values the influence of less-well 
understood soft processes is reduced.\\
The jet photon -- proton cross section, $d\sigma_{\gamma p}$, can be written as
$$
 d\sigma_{\gamma p} = \sum_{ab} \int_{x_\gamma} \int_{x_p} dx_p dx_\gamma
                      f_p(x_p,\mu^2) f_\gamma(x_\gamma,\mu^2)
                  d\hat{\sigma}_{ab}(x_p,x_\gamma,\mu^2)\cdot(1+\delta_{hadr})
$$
where $f_p$ is the proton parton density function (PDF), $f_\gamma$ is the 
photon PDF, $\hat{\sigma}_{ab}$ describes the hard partonic cross section, 
$\mu$ represents both the factorisation and renormalisation scales, $x_\gamma$
is the fraction of the photon's energy participating in the generation of jets 
and $x_p$ is the fractional momentum at which the partons inside the proton are
probed. The hadronisation correction, $\delta_{hadr}$, takes into account 
non-perturbative effects. It can be estimated using Monte Carlo models for the 
parton cascade and fragmentation. For the direct component the photon PDF 
reduces to the Dirac $\delta$-function at $x_\gamma = 1$.\\
The cross sections for the inclusive jet photoproduction were measured by the 
H1 \cite{h1incljet} and ZEUS \cite{zeusincljet} collaborations. The ZEUS 
measurement is presented in Figure \ref{fig:zeusinclusivejet}. The LO QCD 
calculations fail to reproduce the data. 
The next-to-leading order QCD (NLO QCD) predictions deliver a good description 
of the measured distribution. It was found \cite{h1incljet} that the cross 
section calculated with the GRV \cite{grvg} photon PDF gives values which are 
5-10\% larger than those obtained with AFG \cite{afgho}. Different 
parameterisations of the proton PDF have a small effect at low values of 
$E_T^{jet}$. With increasing jet transverse energy differences appear when 
CTEQ5M \cite{cteq5m} based calculations are compared to those obtained with 
MRST99 \cite{mrst99} or CTEQ5HJ \cite{cteq5m}.
\\
The ZEUS collaboration measured the scaled invariant cross section,\\
$E^{jet}_T)^4 (E^{jet}d^3\sigma/dp^{jet}_Xp^{jet}_Yp^{jet}_Z$ where
$E^{jet}_T$ is the jet tranverse energy and $E^{jet}$ is the jet energy. 
The measurement was performed for jets with the pseudorapidity 
$-2< \eta_{\gamma p}^{jet}<0$ measured in the 
photon--proton centre-of-mass frame at the two values of photon--proton 
centre-of-mass energy $W = 180$ GeV and $W = 255$ GeV. The ratio of the scaled 
invariant cross sections when plotted as a function of variable 
$x_T = 2E^{jet}_T/W$ shows the scaling violation. This is depicted in Fig. 
\ref{fig:zeusscaling}.\\
The inclusive jet cross section can be used to determine the value of the 
strong coupling constant, $\alpha_s(M_Z)$. The measured value
$$\alpha_s(M_Z) = 0.1224 \pm 0.0001(stat.)^{+0.0022}_{-0.0019}(exp.)
                                          ^{+0.0054}_{-0.042}(th.)
$$
is consistent with the world average $\alpha_s(M_Z) = 0.1183\pm0.0027$ 
\cite{bethke} (see Fig. \ref{fig:zeusalfas}a) and the measurements 
\cite{zeusalpha,h1alpha} in NC DIS and the $p\bar{p}$ interactions 
\cite{ppalpha}. When plotted as a function of the jet 
transverse energy $\alpha_s$ shows (see Figs. \ref{fig:zeusalfas}b and 
\ref{fig:zeusalfas}c) a clear running behaviour.

\begin{figure}[th]
\begin{center}
\epsfig{file=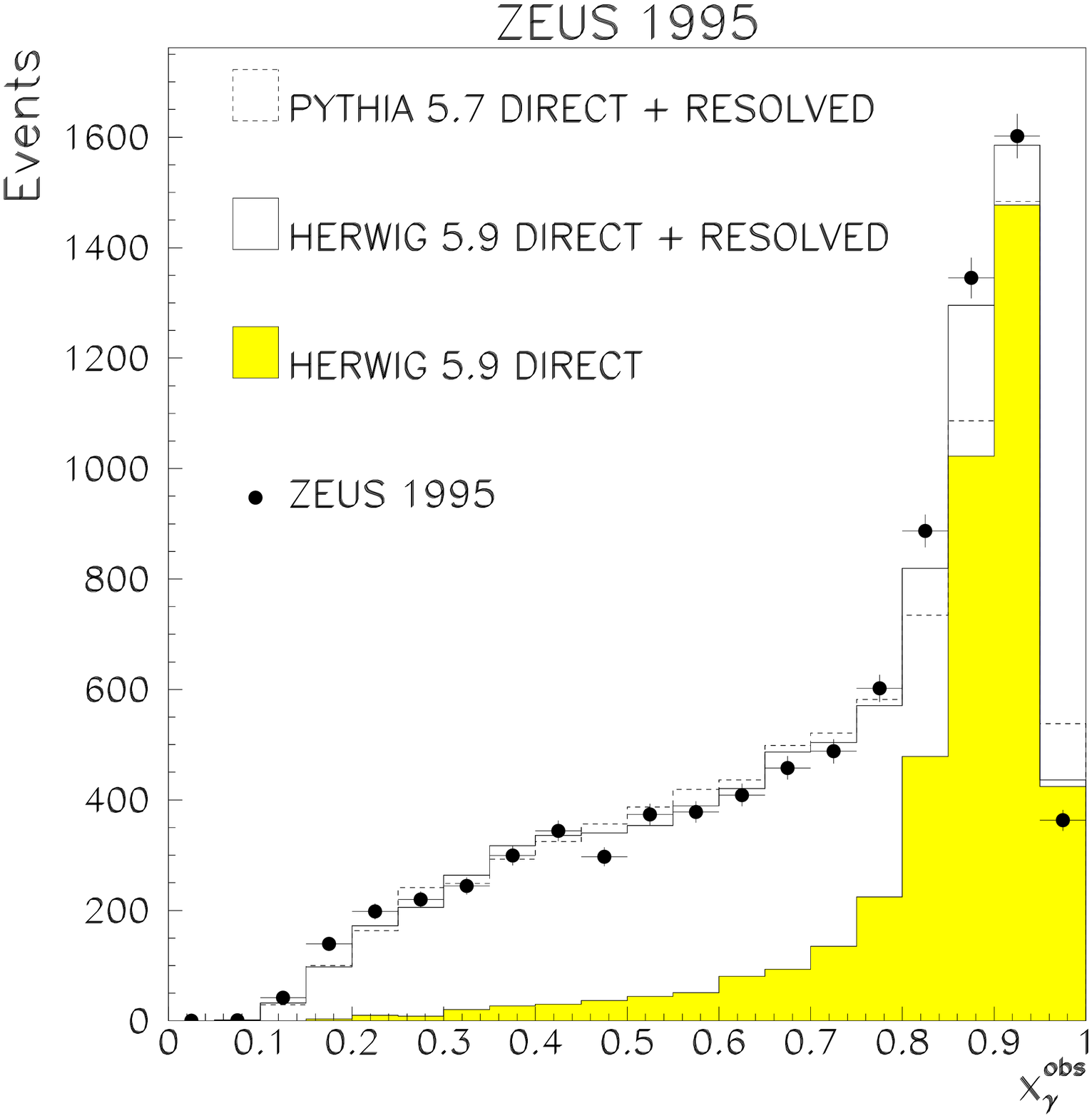,width=0.7\textwidth}
\caption{The $x_\gamma^{obs}$ distribution for the data \cite{zeushetdij} 
compared to MC predictions. The simulated distributions were fitted to the 
data. From \cite{zeushetdij}.}
\label{fig:xgamma}
\end{center}
\end{figure}

\begin{figure}[ht]
\begin{center}
\epsfig{file=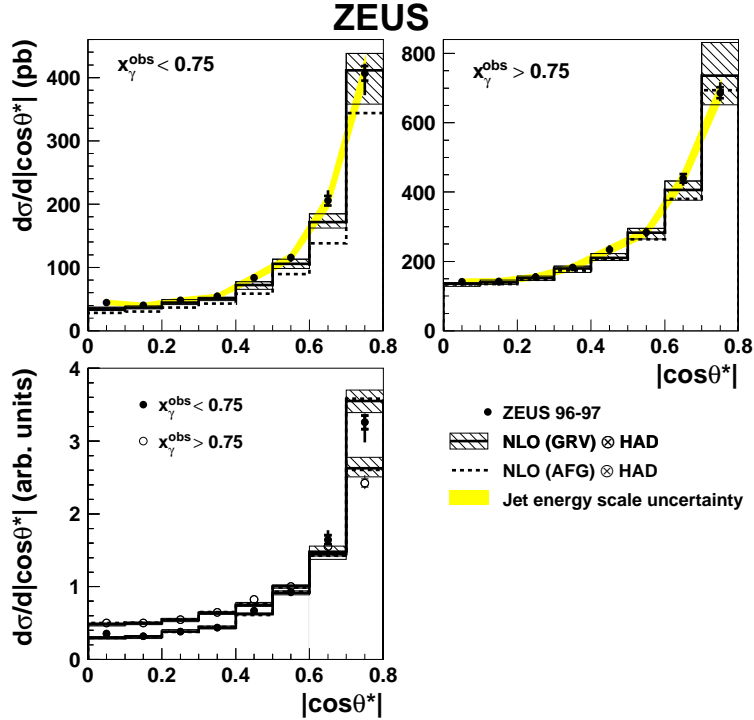,width=10cm}
\end{center}
\caption{Measured cross sections as a function of $|cos\Theta^\star|$ for
$x^{obs}_\gamma < 0.75$ (a) and $x^{obs}_\gamma > 0.75$ compared to NLO 
predictions obtained using GRV-HO and CTEQ5M1 PDFs for the photon and proton 
respectively. Hatched band represents theoretical uncertainties. Shaded band 
shows the jet energy uncertainty. Predictions using AFG-HO are depicted as 
dashed line. In (c) the cross sections are area normalised and the data for
$x^{obs}_\gamma < 0.75$ (solid circles) and for $x^{obs}_\gamma > 0.75$ (open
circles) are shown. From \cite{zeusdij}.}
\label{fig:dijtheta}  
\end{figure}

For the dijet photoproduction $x_\gamma$ is estimated by $x_\gamma^{obs}$ 
which measures the fraction of the photon energy participating in the 
production of the two highest energy jets \cite{xobsdef}
$$
x_\gamma^{obs} = \frac{E_T^{jet1}e^{-\eta_{jet1}}+E_T^{jet2}e^{-\eta_{jet2}}}
                      {2~y~E_e}
$$
\clearpage
\noindent
where $E_T^{jet1,2}$ are the transverse energies of the jets in the laboratory
frame, $\eta_{jet1,2}$ are the jets' pseudorapidities and $y$ is the fraction
of the incident lepton energy carried by the photon in the proton rest frame. 
In leading order QCD $x_\gamma = x_\gamma^{obs}$. The distribution of 
$x_\gamma^{obs}$ is shown in Fig. \ref{fig:xgamma} together with Pythia 
\cite{pythia} and Herwig \cite{herwig} Monte Carlo predictions. The resolved 
component dominates below $x_\gamma^{obs} \approx 0.8$ while above this value 
the direct processes are more important \cite{h1dij,zeusdij}.\\
The distribution of the angle, $\Theta^\star$, between the jets in the
parton--parton cms can be used to test the dynamics of the dijet 
photoproduction. For two-to-two massless parton scattering
$$
cos\Theta^\star = tanh\left(\frac{\eta^{jet1}-\eta^{jet2}}{2}\right).
$$
\clearpage
\noindent
QCD predicts different dijet angular distributions for the resolved and 
direct components. For the latter, mediated mainly by a quark, the 
distribution is $d\sigma/d|cos\Theta^\star| \sim (1-|cos\Theta^\star|)^{-2}$.
If the process is mediated by a gluon exchange, like in the case of the 
resolved component the distribution is 
$d\sigma/d|cos\Theta^\star| \sim (1-|cos\Theta^\star|)^{-1}$. The dijet 
photoproduction cross sections were measured \cite{h1dij,zeusdij} by both HERA 
collaborations. The ZEUS measurement \cite{zeusdij} of  
$d\sigma/d|cos\Theta^\star|$ is presented in Fig. \ref{fig:dijtheta}. For 
$x^{obs}_\gamma < 0.75$, the region enriched in the resolved component, the 
measure cross section lies above the NLO QCD predictions using GRV-HO for the 
photon PDF. Given the theoretical and experimental uncertainties the NLO 
calculations \cite{nlodij} reasonably well describe the data. The calculations 
using AFG-HO are below that of the GRV-HO. For $x^{obs}_\gamma > 0.75$, the
direct region, the NLO predictions are in agreement with the 
measured cross section. In Fig. \ref{fig:dijtheta}c the shapes of the data and 
the NLO distributions are compared. The data for $x^{obs}_\gamma < 0.75$ rise 
more rapidly with $|cos\Theta^\star|$ than those in the direct component 
dominated region. This is consistent with a difference in the dominant 
propagators. A similar observation was made in \cite{h1dij}.
The agreement between the data and the NLO QCD calculations at high
$x_\gamma^{obs}$ and high transverse energy, where the dependence on the photon
structure is small, show a consistency between the data and the gluon 
distribution in the proton extracted from DIS data. 
Further discrimination between the photon PDFs is difficult due to
large uncertainties in the theory at low transverse energies and both
the theoretical and experimental uncertainties at higher transverse energies.
Further constraints of the parton densities in the photon can be made more 
stringent by including the higher-order or re-summed calculations. 
%\clearpage
%This is
%especially important in the low transverse energy region where the theory has 
%large uncertainties and at high transverse energies wherewhere the NLO QCD calculation have large
%between the current PDFs 
%The ZEUS collaboration studied  \cite{zeusdij} the agreement between the data
%and the NLO QCD calculation as a function of the cuts on the jets transverse 
%energies. This study shows that for a given requirement on $E_T^{jet1}$ 
%a good agreement can be achieved by adjusting the cut on $E_T^{jet2}$ 
% the second jet transverse energy which makes the for $25 < E_T^{jet1} < 35$ GeV and 
%$E_T^{jet2} < 21$ the agreement is good for all values of $x_\gamma^{obs}$.
%F+++++++++++++++++++or larger values of $E_T^{jet2}$ the NLO QCD calculations predict the cross 
%section larger than the measured one  
%agreemento between the data and the NLO QCD calculations is good for
%$x^{obs}_\gamma > 0.75$. f the influence
%of a particular choice of the cuts on the jets transverse energiesIt was also 
%observed \cite{zeusdij} that the NLO predictions and the measured cross 
%sections as a function of $x^{obs}_\gamma$ are consistent within given 
%uncertainties. However, the calculations do not reproduce well the data 
%especially for $x^{obs}_\gamma < 0.8$.  This was connected \cite{zeusdij} to 
%the omission of higher-order contributions in the calculations.

%
\section{Inelastic Photoproduction of $J/\Psi$}
\label{sec:inelpsi}
The inelastic $J/\Psi$ photoproduction arises from direct or resolved 
photon interactions. In perturbative QCD it can be calculated in the 
colour--singlet (CS) and colour--octet (CO) frameworks. In the former case a 
colourless $c\bar c$ pair produced by the hard sub-process is identified with 
the physical $J/\Psi$ meson whereas in the latter, the $c\bar c$ pair is 
produced with non-zero colour and then emits one ore more gluons becoming 
finally a colourless meson. The predictions of the CS model underestimate the 
observed $J/\Psi$ production in $p\bar p$ interactions by a large factor 
\cite{cdf} and this difference can be accounted for by the CO contribution. 
%
%  figures....... 
%
\begin{figure}[ht]
\vspace*{-0.5cm}
\begin{center}
\epsfig{file=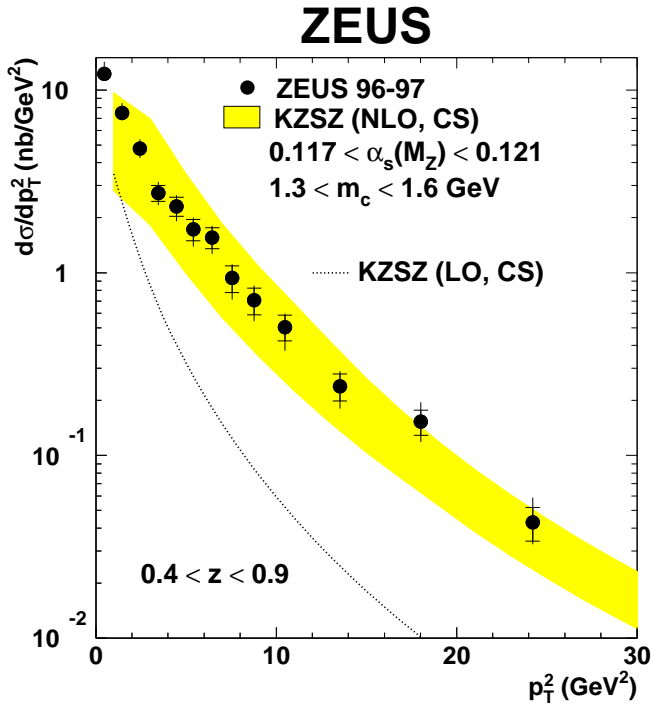,width=0.7\textwidth}
\end{center}
\vspace*{-0.5cm}
\caption{The  $J/\psi$ differential cross-section $d\sigma/dp^2_{T}$. 
The data are compared with two predictions of the colour-singlet model  
described in the text. From \cite{inelpsi}.
\label{fig:inelpt} }
\end{figure}
\begin{figure}[ht]
\vspace*{-0.5cm}
\begin{center}
\epsfig{file=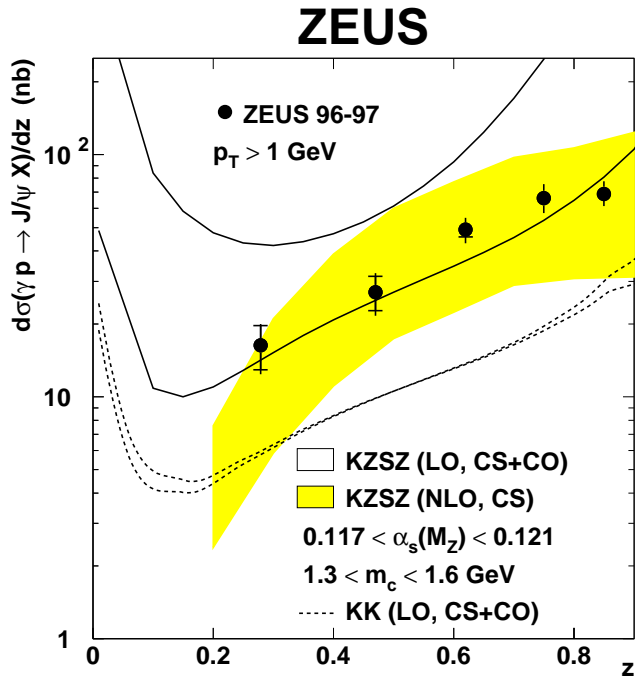,width=0.7\textwidth}
\end{center}
\vspace*{-0.5cm}
\caption{The  $J/\psi$ differential cross-section $d\sigma/dz$ for $p_T > 1$ 
GeV. The data are compared with predictions of the colour-singlet model and two
predictions including both the colour-singlet and colour-octet contributions 
described in the text (from \cite{inelpsi}).
\label{fig:inelz} }
\end{figure}
\\
The ZEUS collaboration investigated \cite{inelpsi} the inelastic charmonium 
($J/\Psi$ and $\Psi^\prime$ ) photoproduction in the energy range 
$50 < W < 180$ GeV, through their decays into muon pairs. The $J/\Psi$ 
production cross section as a function of its transverse momentum, $p_T$, and 
the inelasticity, z (the fraction of incoming photon's energy carried by the 
$J/\Psi$) is shown in Figs. \ref{fig:inelpt} and \ref{fig:inelz} and are 
compared with theoretical calculations mentioned previously. A prediction of 
the colour--singlet model in the leading logarithms approximation (LO, CS) 
clearly does not describe the $p_T$ distribution. Including next-to-leading 
corrections (NLO, CS) it matches the data very well suffering however from some
theoretical uncertainty \cite{KZSZ}. The same is valid for the inelasticity 
distribution -- see Fig. \ref{fig:inelz}. In this figure also the predictions 
of two particular calculations using both singlet and octet colour mechanisms 
and the leading logarithms approximation (LO, CS+CO) are presented. The NLO QCD
calculations provide a prediction which is consistent with the data within 
large uncertainties resulting from extracting the CO matrix elements 
\cite{KZSZ, KK}. These inconclusive results mean that a quantitative 
understanding of the $J/\Psi$ production mechanism is still lacking.

\section{Beauty Photoproduction}
\label{sec:beauty}
Beauty photoproduction, owing to the large mass of the $b$ quark which provides
a hard scale, is a stringent test of perturbative QCD. 
\begin{figure}[ht]
\vspace*{-0.5cm}
\begin{center}
\epsfig{file=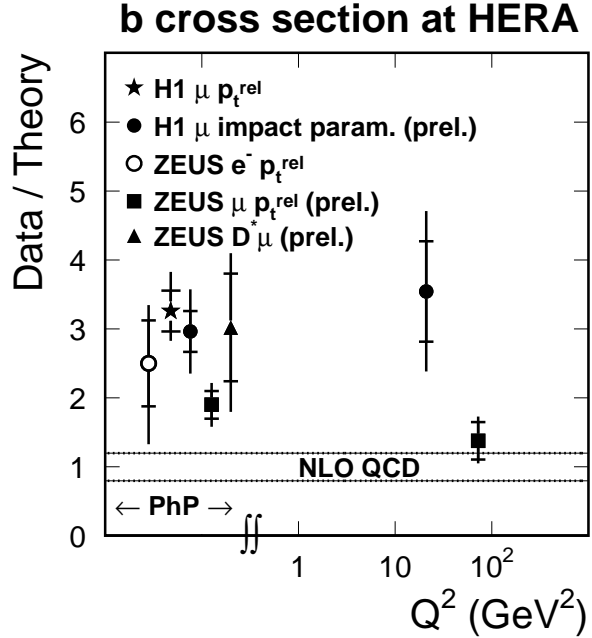,width=0.75\textwidth}
\end{center}
\vspace*{-0.5cm}
\caption{Ratio of the measured $b$-production cross section at HERA and the 
theoretical expectation from NLO QCD, as a function of $Q^2$. A star and 
circles represent older H1 and ZEUS results \cite{b-old}. A triangle represents the photoproduction measurement using $D^\star+\mu$ tag \cite{d-b-mu}.
From \cite{zeus-bphp}.
\label{fig:b-php} }
\end{figure}
The ZEUS collaboration investigated  \cite{zeus-bphp} this process using events
with two high transverse energy jets and a muon in the final state. The 
fraction of beauty quarks in the data was determined using the transverse 
momentum distribution of the muon relative to the closest jet. The total 
and differential cross sections for the process 
$ep \rightarrow  b\bar{b} \rightarrow 2 ~jets + X $ were determined using 
Monte Carlo models to extrapolate for the unmeasured part of the muon 
kinematics and to correct for the inclusive branching ratio 
$B(b \rightarrow \mu )$. The measured cross sections were compared to NLO QCD 
predictions based on the program by Frixione et al. \cite{Frixione}. This is 
summarised in Fig. \ref{fig:b-php} where the ratio of the measured to the 
predicted cross section is presented as a function of $Q^2$. For $Q^2 \sim 0$ 
this ratio is about 2 which demonstrates that the model considerably 
underestimates the beauty photoproduction. The differential cross section in 
the region of good muon acceptance is also larger than the theoretical 
prediction however compatible with it within the experimental and theoretical 
uncertainties. The excess of $b$-quark production over NLO QCD predictions was 
also found in $p\bar{p} $ annihilations (see references in \cite{zeus-bphp}). 
The above observations are a challenge for the perturbative QCD.

\section{Summary}
\label{sec:sum}
%%%%%%%%%%%%%%%
%
% summary
%
%%%%%%%%%%%%%%%

Selected aspects of the photoproduction study with the ZEUS detector at HERA 
have been presented. 
%In all of them the hadronic nature of the photon is demonstrated. 
The photon--proton interactions show many features similar to soft 
hadron--hadron collisions. However, in the  presence of a large scale, 
delivered by the meson mass or the transverse energy, the hard scattering of 
partonic constituents in the photon and proton becomes important. Many 
particular features of the hard $\gamma p$ interactions are successfully 
described by perturbative QCD based models.

\section*{Acknowledgments}
%%%%%%%%%%%%%%%
%
%        acknowledgements
%
%%%%%%%%%%%%%%%

We gratefully acknowledge support of the DESY directorate during our stays 
at DESY. We thank our colleges from the ZEUS collaboration for their help,
co-operation and creation of a stimulating atmosphere.
The authors thank prof. E. Lohrmann for discussions and critical reading
of the manuscript.


\clearpage
\begin{thebibliography}{999} 
\addcontentsline{toc}{section}{References}
   %%%%%%%%%%%%%%%%%%%%%%%
%
% intro
%
%%%%%%%%%%%%%%%%%%%%%%%
%
\bibitem{bauer}
   T.H. Bauer et al., Rev. Mod. Phys. {\bf 50} (1978) 261.
%
\bibitem{kogan} E. Kogan, Ph D Thesis, Weizmann Institute, Rehovot, Israel (unpublished);\\
See also \cite{bauer} p. 270.
%
\bibitem{pdg} Particle Data Group, K. Hagiwara et al., Phys. Rev.  {\bf D66} (2002) 010001.
%
\bibitem{alven} H. Alvensleben et al., Phys. Rev. Lett. {\bf 30} (1973) 328.
%
\bibitem{ioffe} B. L. Ioffe, Phys. Lett. {\bf B30} (1969) 123;\\
 B. L. Ioffe, V. A. Khoze and L. N. Lipatov, Hard Proccesses, vol 1 (1984) 155.
\bibitem{frisjo} G. A. Schuler and T. Sj\"ostrand, Nucl. Phys. {\bf B407} (1993) 529;\\
 G. A. Schuler and T. Sj\"ostrand, CERN-TH 6796/93.
%
\bibitem{vdm} J. J. Sakurai, Ann. Phys. (NY) {\bf 11} (1960) 1;\\
              J. J. Sakurai, Phys. Rev. Lett. {\bf 22} (1969) 981.
%%%%%%%%%%%%%%%%%%%%%%%%%%%%%%%%
%
% experimental env.
%
%%%%%%%%%%%%%%%%%%%%%%%%%%%%%%%
%
\bibitem{hera} 
   A Proposal for a Large Electron - Proton Colliding Beam 
   Facility at DESY, DESY HERA 81-10 (1981);\\
   B. H. Wiik, Electron - Proton Colliding Beams, The Physics Programme and 
   the Machine, Proc. 10$^{th}$ SLAC Summer Institute, ed. A. Mosher, (1982) 
   233.
%
\bibitem{zeusblue}
   ZEUS Collaboration, The ZEUS Detector, Technical Proposal, (1986);\\
   The ZEUS Detector, Status Report 1993, ed. U. Holm, DESY (1993);\\
   http://www-zeus.desy.de/bluebook/bluebook.html.
%
\bibitem{ctd}
   B. Foster et al., NIM {\bf A338} (1993) 254.
%   B. Foster et al., Nucl. Instrm. Meth. {\bf A338} (1993) 254.
%
\bibitem{ucal}
   M. Derrick et al., NIM {\bf A309} (1993) 77;\\
   A. Berstein et al., NIM {\bf A336} (1993) 23;\\
   A. Anderson et al., NIM {\bf A309} (1991) 101.
%   M. Derrick et al., Nucl. Instrm. Meth. {\bf A309} (1993) 77;\\
%   A. Berstein et al., Nucl. Instrm. Meth. {\bf A336} (1993) 23;\\
%   A. Anderson et al., Nucl. Instrm. Meth. {\bf A309} (1991) 101.
%
\bibitem{lumiexp} 
   J.~Andruszk\'ow et al., DESY 92-066 (1992);\\
   J.~Andruszk\'ow et al., Acta Phys. Polon. {\bf B32} (2001) 2025.
%%%%%%%%%%%%%%%%%%%%%%%%%%%%
%
% sigtot
%
%%%%%%%%%%%%%%%%%%%%%%%%%%%
%
\bibitem{zeussigtot} ZEUS Collab., S. Chekanov et al., Nucl. Phys. {\bf B627}
(2002) 3.
%
\bibitem{h1sigtot}
  H1 Collab., S. Aid et al., Z. Phys. {\bf C69} (1995) 27.
%
\bibitem{bpc} 
  ZEUS Collab., M. Derrick et al., Eur. Phys. J. {\bf C7} (1999) 609.
%
\bibitem{bpcgvdm} 
  J. J. Sakurai and D. Schildknecht, Phys. Lett. {\bf B40} (1972) 121.
%
\bibitem{regge} T. Regge, Nuovo Cimento {\bf 14} (1959) 951;\\
                T. Regge, Nuovo Cimento {\bf 18} (1960) 947.
%
\bibitem{collins}
  P.D.B.~Collins, An Introduction to Regge Theory and High Energy 
   Physics, Cambridge University Press (1977).
%
\bibitem{dl98}
  A.~Donnachie and P.V.~Landshoff, Phys.~Lett. {\bf B437} (1998) 408.
%
\bibitem{cudell} J. R. Cudell et al, Phys. Rev. {\bf D61} (2000) 034019,
                 Erratum-ibid. {\bf D63} (2001) 059901.
%
\bibitem{compass} HERA and Compass Groups, S. I. Alekhin et al., CERN-HERA-8701 (1987).
%
\bibitem{frisjo2}
C. Friberg and T. Sj\"ostrand, J. High Energy Phys. {\bf 09} (2000) 10.
%
\bibitem{block}
  M.M.~Block et al., Phys.~Rev. {\bf D60} (1999) 054024;\\
  M.M.~Block, F.~Halzen and T.~Stanev, Phys.~Rev. {\bf D62} (2000) 077501.
%
\bibitem{allm97} 
   H. Abramowicz, E. Levy, DESY-97-251.
%   H.~Abramowicz et al., Phys. Lett. {\bf B269} (1991) 465.
%  H.~Abramowicz, E.~Levin, A.~Levy and U.~Maor, Phys. Lett. {\bf B269} (1991) 465.
%
\bibitem{l3fact} L3 Collab., M. Acciarri et al., Phys. Lett. {\bf B519} (2001) 33.
%
\bibitem{opalfact} OPAL Collab., G. Abbiendi et al., Eur. Phys. J. {\bf C14} (2000) 199.
%%%%%%%%%%%%%%%%%%%%%%%
%
% elastic
%
%%%%%%%%%%%%%%%%%%%%%%
%
%
\bibitem{dl92}
A.~Donnachie and P. V.~Landshoff, Phys. Lett. {\bf B296} (1992) 227.
%
\bibitem{RHOMEAS}
   ZEUS Collab., J.~Breitweg et al., Eur. Phys. J. {\bf C2} (1998) 247.
%
\bibitem{OMEGAMEAS}
   ZEUS Collab., M.~Derrick et al., Z. Phys. {\bf C73} (1996) 73.
%
\bibitem{PHIMEAS}
   ZEUS Collab., M.~Derrick et al., Phys. Lett. {\bf B377} (1996) 259.
%
\bibitem{h1rho} H1 Collab., S. Aid et al., Nucl. Phys. {\bf B463} (1996) 3.
%
\bibitem{herabvsw}
ZEUS Collab., M. Derrick et al., Z. Phys. {\bf C69} (1995) 39.
%
\bibitem{lowbvsw}
W. G. Jones et al., Phys. Rev. Lett. {\bf 21} (1968) 586;\\
C. Berger et al., Phys. Lett. {\bf B39} (1972) 659;\\
SBT Collab. J. Ballam et al. Phys. Rev. {\bf D5} (1972) 545;\\
G. E. Gladding et al., Phys. Rev. {\bf D8} (1973) 3721.
%
\bibitem{vmvsw} P. Fleischmann for H1 Collab., to appear in
                Proc. of DIS 2003, 23--27 April, St. Petersburg, Russia.
%
%\bibitem{h1psi} H1 Collab., S. Aid et al., Nucl. Phys {\bf B472} (1996) 3.
\bibitem{h1psi1} H1 Collab., C. Aldoff et al., Phys. Lett. {\bf B483} (2000) 23.
%
\bibitem{h1psi2} H1 Collab., Paper submitted to XXI Int. Europphysics 
Conference on High Energy Physcis, July 17-23, Aachen, abstract 108.
%
\bibitem{zeuspsi} ZEUS Collab., S. Chekanov et al.,  Eur. Phys. J. {\bf C24} (2002) 345.
%
\bibitem{ryskin} M. Ryskin, Z. Phys. {\bf C57} (1993) 89.
%
%\bibitem{ryskin2} M. Ryskin et al., Z. Phys. {\bf C76} (1997) 231.
%
\bibitem{zeusupsi} ZEUS Collab., J. Breitweg et al., Phys. Lett. {\bf B437} (1998) 432.
%
\bibitem{frankfurtupsi} L. Frankfurt, M. McDermott and M. Strickman, 
J. High Energy Phys. {\bf 02} (1999) 002.
%
\bibitem{martinupsi}  
A. D. Martin, M. G. Ryskin and T. Teubner, Phys. Lett. {\bf B454} (1999) 339.
%%%%%%%%%%%%%%%%%%%%%%%%
%
% diffractive
%
%%%%%%%%%%%%%%%%%%%%%%%%
%
\bibitem{e612} T. J. Chapin et al., Phys. Rev. {\bf D31} (1985) 17.
%
\bibitem{h1gdiss} H1 Collab., C. Adloff et al., Z. Phys. {\bf C74} (1997) 221.
%
\bibitem{dissfrac}
   ZEUS Collab., J.~Breitweg et al., Z. Phys. {\bf C75} (1997) 421.
%
%\bibitem{h1lrg}  H1 Collab., T. Ahmed et al., Nucl. Phys. {\bf B429} (1994) 477.
%
%\bibitem{zeuslrg} ZEUS Collab., M. Derrick et al., Phys. Lett. {\bf B315} (1993) 481.
%
\bibitem{islrg} G. Ingelmann and P. E. Schlein, Phys. Lett. {\bf B152} (1985) 256.
%
\bibitem{bjorkenlrg} J. D. Bjorken, Phys. Rev. {\bf D47} (1993) 101.
%
\bibitem{zeusjetlrg} ZEUS Collab., M. Derrick et al., Phys. Lett. {\bf B356} (1995) 129.
%
\bibitem{h1jetlrg} H1 Collab., T. Ahmed et al., Nucl. Phys. {\bf B435} (1995) 3.
%
\bibitem{pythia} 
T. Sj\"ostrand, Computer Phys. Commun. {\bf 82} (1994) 74;\\
T. Sj\"ostrand, PYTHIA 5.7, hep-ph-/9508391.
%
\bibitem{h1diff2j} H1 Collab., C. Adloff et  at., Eur. Phys. J. {\bf C6} (1999) 421.
%
\bibitem{zeusdiff2j} ZEUS Collab., J. Breitweg et al., Eur. Phys. J. {\bf C5} (1998) 45.
%
\bibitem{h1diff2jb} H1 Collab., contribution to International Europhysics Conference on High Energy Physics, EPS03, July 17-23, 2003, Aachen, abstract 087.
%
\bibitem{zeusjlrgj} ZEUS Collab., M. Derrick et al., Phys. Lett. {\bf B369} (1996) 55.
%
\bibitem{h1jlrgj} H1 Collab., C. Adloff et al., Eur. Phys. J. {\bf C24} (2002) 517.
%
\bibitem{cdfgap} CDF Collab., F. Abe et al., Phys. Rev. Lett. {\bf 74} (1995) 855.
%
\bibitem{d0gap} D0 Collab., S. Abachi et al., Phys. Rev. Lett. {\bf 76} (1996) 734.
%
\bibitem{h1diff} H1 Collab., C. Aldloff et al., Z. Phys. {\bf C76} (1997) 613.
%
\bibitem{zeusdiff} ZEUS Collab., J. Breitweg et al., Eur. Phys. J. {\bf C6} (1999) 43.
%
% zmiana  JF, 19.09-------------------------------------------------------
\bibitem{zeusvmt2} ZEUS Collab., S. Chekanov et al., Eur. Phys. J. {\bf C26} (2003) 389.
%
% zmiana JF--------------------------------------------------------------
\bibitem{polud} J. R. Forshaw and G. Poludniowski, %preprint hep-ph/0107068, 
Eur. Phys. J. {\bf C26} (2003) 411.
%
\bibitem{pqcdvmt} J. Bartels et al., Phys. Lett {\bf B375} (1996) 301;\\
                  D. Yu. Ivanov et al., Phys. Lett {\bf B478} (2000) 101,
                  Erratum-ibid. {\bf B348} (2001) 295.
%
\bibitem{zeusvmt1} ZEUS Collab., J. Breitweg et al., Eur. Phys. J. {\bf C14} (2000) 213.
%
% dodane , JF, 19.09------------------------------------------------------
\bibitem{Enberg} R. Enberg, L. Motyka and G. Poludniowski, 
Eur. Phys. J. {\bf C26} (2003) 219.
%
\bibitem{EnbergAPP} R. Enberg, L. Motyka and G. Poludniowski, %preprint hep-ph/0207034, 
Acta Phys. Pol. {\bf B33} (2002) 3511.
%
\bibitem{H103} H1 Collab., A. Aktas et al., DESY 03-061 (June 2003).
%
%%%%%%%%%%%%%%%%%%%%%%%%
%
% hard
%
%%%%%%%%%%%%%%%%%%%%%%%
%
\bibitem{h1incljet} H1 Collab.,  C. Adloff et al., Eur. Phys. J. {\bf C29} (2003) 497.
%
\bibitem{zeusincljet} ZEUS Collab., S. Chekanov et al., Phys. Lett. {\bf B560} (2003) 7.
%
\bibitem{grvg} 
   M.~Gl\"uck, E.~Reya and  A.~Vogt, Phys. Rev., {\bf D45} (1992) 3986;\\
   M.~Gl\"uck, E.~Reya and  A.~Vogt, Phys. Rev., {\bf D46} (1992) 1973.
%
\bibitem{afgho} P. Aurenche, J. P. Guillet and M. Fontanaz, Z. Phys. {\bf C64} (1994) 621.
%
\bibitem{cteq5m} CTEQ Collab., H. L. Lai et al., Eur. Phys. J. {\bf C12} (2000) 375.
%
%\bibitem{mrst99} A. D. Martin, R. G. Roberts, W. J. Stirling and R. S. Thorne, 
\bibitem{mrst99} A. D. Martin et al., Eur. Phys. J. {\bf C14} (2000) 133.
%
\bibitem{bethke} S. Bethke, J. Phys. {\bf G26} (2000) R27;\\
                 S. Bethke, hep-ex/0211012.
%
\bibitem{zeusalpha} ZEUS Collab., J. Breitweg et al., Phys. Lett. {\bf B507} (2001) 70;\\
                  ZEUS Collab., S. Chekanov et al., Phys. Lett. {\bf B547} (2002) 164.
%
\bibitem{h1alpha} H1 Collab., C. Adloff et al., Eur. Phys. J. {\bf C19} (2001) 289.
%
\bibitem{ppalpha} CDF Collab., T. Affolder et al., Phys. Rev. Lett. {\bf 88} (2002) 042001.
%
\bibitem{xobsdef} ZEUS Collab., M. Derrick et al., Phys. Lett. {\bf B348} (1995) 665.
%
%\bibitem{h1jets}  H1 Collab., S. Aid et al., Z. Phys. {\bf C70} (1996) 17.
%
\bibitem{h1dij} H1 Collab., C. Adloff et al., Eur. Phys. J. {\bf C25} (2002) 13.
%
\bibitem{zeusdij} ZEUS Collab., S. Chekanov et al., Eur. Phys. J. {\bf C23} (2002) 615.
%
\bibitem{zeushetdij} ZEUS Collab., J. Breitweg et al., Eur. Phys. J. {\bf C11} (1999) 35.
%
\bibitem{herwig} 
G. Marchesini and B. R. Webber, Nucl. Phys. {\bf B238} (1984) 1;\\
B. R. Weber, Nucl. Phys. {\bf B238} (1984) 492;\\
G. Marchesini and B. R. Webber, Nucl. Phys. {\bf B310} (1988) 461;\\
G. Marchesini et al., Computer Phys. Commun. {\bf 67} (1992) 465;\\
G. Marchesini et al., HERWIG 5.9, hep-ph/9607393.
%
\bibitem{nlodij} S. Frixione and G. Ridolfi, Nucl. Phys. {\bf B507} (1997) 4007;\\
                 S. Frixione, Nucl. Phys. {\bf B507} (1997) 295.
%
%\bibitem{heisenberg} W. Heisenberg, Z. Phys. \vol(133) (1952) 65.
%
%%%%%%%%%%%%%%%%%%%%%%%%%%%%%%
%
%   inelstic charmonium
%
%%%%%%%%%%%%%%%%%%%%%%%%%%%%%%
%
\bibitem{inelpsi} ZEUS Collab., S. Chekanov et al., Eur. Phys. J. {\bf C27} (2002) 173.
\bibitem{cdf} CDF Collab., F. Abe et al, Phys. Rev. Lett. {\bf 79} (1997) 572;
\\ CDF Collab., F. Abe et al, Phys. Rev. Lett. {\bf 79} (1997) 578.
\bibitem{KZSZ} M. Kr\"amer et al, Phys. Lett. {\bf B348} (1995) 657;
\\ M. Kr\"amer et al, Nucl. Phys. {\bf B459} (1996) 3;
\\ M. Kr\"amer, Prog. Part. Nucl. Phys., {\bf 47} (2001) 141.
\bibitem{KK} B. A. Kniehl and G. Kramer, Eur. Phys. J. {\bf C6} (1999) 493.
%
%%%%%%%%%%%%%%%%%%%%%%%%%%
%
%     b production
%
%%%%%%%%%%%%%%%%%%%%%%%%%%%
%
\bibitem{zeus-bphp} ZEUS Collab., contribution to the XXXIst International 
Conference on High Energy Physics, 24-31 July 2002, Amsterdam, abstract 785.
\bibitem{Frixione} S. Frixione et al., Phys. Lett. {\bf B348} (1995) 633.
\bibitem{b-old} H1 Collab., C. Adloff et al., Phys. Lett. {\bf B467} (1999) 156;\\
ZEUS Collab., J. Breitweg et al., Eur. Phys. J. {\bf C18} (2001) 625;\\
H1 Collab., contribution to  XXXth International Conference on High Energy 
Physics, July 12-18, 2001, Budapest, Hungary, abstract 807.
\bibitem{d-b-mu} ZEUS Collab., contribution to the XXXIst International 
Conference on High Energy Physics, 24-31 July 2002, Amsterdam, abstract 784.

\end{thebibliography}
\end{document}